# Control of magnetic anisotropy by orbital hybridization in $(La_{0.67}Sr_{0.33}MnO_3)_n/(SrTiO_3)_n$ superlattice

Bangmin Zhang,[1,†] Lijun Wu,[2,†] Jincheng Zheng,[2,3] Ping Yang,[4] Xiaojiang Yu,[4] Jun Ding,[1] Steve M. Heald,[5] R. A. Rosenberg,[5] T. Venkatesan,[1,6,7,8,9] Jingsheng Chen,[1] Cheng-Jun Sun,[5,*] Yimei Zhu,[2,*] and Gan Moog Chow[1,*]

[1]Department of Materials Science & Engineering, National University of Singapore, 9 Engineering Drive 1, 117576, Singapore.

[2]Condensed Matter Physics & Materials Science Division, Brookhaven National Laboratory, Upton, New York, 11973, USA.

[3]Department of Physics, Xiamen University, Xiamen 361005, People's Republic of China.

[4]Singapore Synchrotron Light Source (SSLS), National University of Singapore, 5 Research Link, 117603, Singapore.

[5]Advanced Photon Source, Argonne National Laboratory, Argonne, Illinois, 60439, USA.

[6]NUSNNI-Nanocore, National University of Singapore, 2 Engineering Drive 3, 117581, Singapore.

[7]Department of Physics, National University of Singapore, 2 Science Drive 3, 117551, Singapore.

[8]Department of Electrical & Computer Engineering, National University of Singapore, 4 Engineering Drive 3, 117583, Singapore.

[9]NUS Graduate School for Integrative Science and Engineering, National University of Singapore, 28 Medical Drive, 117456, Singapore.

[†]Authors contribute to the work equally.

*Corresponding authors: cjsun@aps.anl.gov, zhu@bnl.gov, msecgm@nus.edu.sg

**Abstract**:

The asymmetry of chemical nature at the hetero-structural interface offers an unique opportunity to design desirable electronic structure by controlling charge transfer and orbital hybridization across the interface. However, the control of hetero-interface remains a daunting task. Here, we report the modulation of interfacial coupling of $(La_{0.67}Sr_{0.33}MnO_3)_n/(SrTiO_3)_n$ superlattices by manipulating the periodic thickness with $n$ unit cells of $SrTiO_3$ and $n$ unit cells $La_{0.67}Sr_{0.33}MnO_3$. The easy axis of magnetic anisotropy rotates from in-plane ($n = 10$) to out-of-plane ($n = 2$) orientation at 150 K. Transmission electron microscopy reveals enlarged tetragonal ratio > 1 with breaking of volume conservation around the $(La_{0.67}Sr_{0.33}MnO_3)_n/(SrTiO_3)_n$ interface, and electronic charge transfer from Mn to Ti $3d$ orbitals across the interface. Orbital hybridization accompanying the charge transfer results in preferred occupancy of $3d_{3z^2-r^2}$ orbital at the interface, which induces a stronger electronic hopping integral along the out-of-plane direction and corresponding out-of-plane magnetic easy axis for $n = 2$. We demonstrate that interfacial orbital hybridization in superlattices of strongly correlated oxides may be a promising approach to tailor electronic and magnetic properties in device applications.

The asymmetry at hetero-structural interface of 3$d$ transitional metal ABO$_3$ oxides, including the mismatch of lattice constant, oxygen octahedral rotation and distortion, and chemical environment, has profound influences on spin and orbital coupling, yielding emerging phenomena, such as enhanced ordering temperature, induced interfacial magnetism at superconductor/manganite interface and orbital reconstruction.[1-5] Coupling between crystal structures[6-12] across the heterostructure interface leads to new properties, for example, film-thickness dependent interfacial ferromagnetic-polaronic insulator in Pr$_{0.67}$Sr$_{0.33}$O$_3$/SrTiO$_3$,[13] spontaneous deep polarization[9] in SrTiO$_3$ near the La$_{2/3}$Sr$_{1/3}$MnO$_3$/SrTiO$_3$ interface, rotation of magnetic easy axis of La$_{2/3}$Sr$_{1/3}$MnO$_3$[14] and SrRuO$_3$[15] thin film. In strongly correlated 3$d$ oxides, the properties of transition metal oxides are sensitive to electronic structure, especially the occupation of 3$d$ orbitals,[4] which may be manipulated by doping with different covalent ions,[16] hydraulic pressure,[17] and external electric bias.[18-19] Control of interfacial electronic structure may also be exploited using chemical asymmetry, such as polar discontinuity, to induce charge transfer and orbital hybridization across the interface.[20-22]

La$_{0.67}$Sr$_{0.33}$MnO$_3$ (LSMO) has a pseudo-cubic lattice of 3.880 Å at room temperature, high Curie temperature and high spin polarization.[22] SrTiO$_3$ (STO) is cubic[23] at room temperature with lattice constant ~ 3.905 Å close to that of LSMO. For (00$l$) orientated heterostructure, the TiO$_2$ and SrO atomic plane of STO are charge-neutral (no planar charge), while MnO$_2$ and La/SrO atomic plane of LSMO have negative and positive planar charge, respectively, as illustrated as blue bar at right bottom of Fig. 1**A**. The polar discontinuity exists around the LSMO/STO heterostructure, and the charge transfer across the interface occurs in order to avoid the polar catastrophe.[24-25] The final distribution of planar charge (charge on each atomic layer) is illustrated as pink bar at right bottom of Fig. 1**A**, and the short-range charge transfer is constrained to several unit cells (UC) within the proximity of the interface,[25-27] as illustrated as the probability of charge transfer at left bottom of Fig. 1**A.** Artificial superlattices of $ABO_3$ with multiple interfaces provide a viable way to manipulate the strength of interfacial coupling and corresponding properties. Although properties of LSMO/STO system, such as insulator-metal transition temperature,[28] magnetoelastic effect,[29] and octahedral rotation[30] have been investigated, to date, the electronic charge transfer and related properties at LSMO/STO interfaces remain largely unexplored. Increased number of LSMO/STO interfaces in superlattices may enhance the probability of short-range charge transfer, offering an unique approach to investigate the effects of charge transfer.

In this letter, we report the manipulation of interfacial electronic properties and corresponding magnetic anisotropy in $[(La_{0.67}Sr_{0.33}MnO_3)_n/(SrTiO_3)_n]_m$ superlattices (SL) with different periodic thickness (*n* UC LSMO and *n* UC STO in one period, referred to SL*n* hereafter). The total thickness was kept at 2n×m ~ 120 UC. The periodic thickness of SL*n* was systematically changed to obtain different numbers of LSMO/STO interfaces, as illustrated in Fig. 1**A**. The experimental details are available in supplementary information. The magnetic anisotropy was measured by a superconducting quantum interference device (SQUID) at 150 K and 10 K. At 150K, the magnetic easy axis of SL10 (in-plane) is clearly different from SL2 (out-of-plane), as shown in Fig. 1**B**-1**C**. The element-specific magnetic information was collected using X-ray magnetic circular dichroism (XMCD). The difference in magnetic easy axis of different samples is further confirmed by angular-dependent Mn $L_{3,2}$ edge XMCD in Fig. **S1.** The contribution from Ti sites to the total magnetic moment is insignificant as calculated by the sum rules[31] from Ti $L_{3,2}$ XMCD (Fig. **S2**). Hence the discussion of magnetic properties in this work will mainly focus on the contribution from Mn sites hereafter.

The effective anisotropy may be viewed as the sum of contributions (H$_{//}$) favoring in-plane orientation and that (H$_\perp$) favoring out-of-plane orientation, as illustrated in the inset of Fig. 1B-1C. The effective anisotropy field[32] may be decoupled as $H_{eff} = H_i + H_b - H_d$, including constant bulk term $H_b$, demagnetization term $H_d$ favoring in-plane direction, and interfacial contribution $H_i$ favoring out-of-plane direction. Compared to the bulk, multiple interfaces in superlattices render $H_d$ and $H_i$ non-trivial. The calculated interfacial contribution $H_i$ (Fig. 1E) increases with decreasing *n*. The positive value of *H* is defined to denote the easy axis normal to the film plane and a negative value means the easy axis in the film plane. The change of the interfacial contribution $H_i$ is responsible for the change of easy axis as a function of periodic thickness of the superlattices. Analysis of the magnetic anisotropy at 10 K (Figs. **S3**-**S4**) shows in-plane magnetic easy axis for all samples. Although the total effective magnetic anisotropy at 10 K does not switch due to the strong demagnetization contribution $H_d$, the interfacial contribution $H_i$ shows a similar trend as that at 150 K. Hence, the effect of periodic thickness on interfacial term $H_i$ exists in a wide temperature range.

The effect of interfacial octahedral rotation on magnetic anisotropy has been reported,[14-15, 33] and current work has measured volume-averaged x-ray half-integer diffraction[34-36] to retrieve the information of MnO$_6$ rotation (Fig. **S5**). There is a significant difference in Mn-O bond angle and bond length in SL10 for the in-plane and out-of-plane directions. In SL2, the bond lengths of both directions are similar, yet the bond angles are different. The larger in-plane bond angle for SL2 (Fig. **S5B**) suggests a higher in-plane hopping integral that should favor in-plane magnetic easy axis instead of the out-of-plane direction as observed in current work. Therefore it is necessary to consider other factors that are responsible for the observed magnetic anisotropy. Then high resolution scanning transmission electron microscopy (STEM) were used to probe the local structural information at the LSMO/STO interfaces for $n = 2, 6, 10$, as shown in Fig. 2. The STO substrate was treated with a $TiO_2$ termination layer before deposition, and an atomically-flat interface between LSMO/STO existed in these samples.

By refining the peak positions, the (001) plane spacing of two successive La/SrO (or SrO) planes along out-of-plane direction (hereafter denoted as $c_{int}$) was obtained, and the substrate is at the left side of image. For SL10 and SL6, the $c_{int}$ shows an oscillating pattern with a maximum value at the left STO/LSMO interface of each LSMO layer, larger than 3.905 Å, and with a minimum at the right LSMO/STO interface of each LSMO layer, which indicates an asymmetric coupling at the two interfaces of each LSMO layer and should be related with the growth process (to be discussed later). The oscillation pattern of $c_{int}$ in the film is further confirmed by geometry phase analysis (GPA)[37-38] from HRTEM image (the second row of Fig. 2**E**). There is basically no change of the in-plane lattice (the third row in Fig. 2**E**) over the whole images except some random changes caused by the image noise. For SL2, there is no obvious pattern of $c_{int}$, but $c_{int}$ of the whole film is slightly larger ($c_{int}/a > 1$) than that of the substrate STO, revealed as the dominated red part of GPA (second raw of Fig. 2**F**). The select area electron diffraction (SAED) is used to separate the contribution of diffraction peak between the film and the substrate in Fig. **S6**. It is obvious that the 002 diffraction peak of SL2 film (blue line) shifts to the left with respect to that of the substrate (red line), confirming that the $c_{int}$ of SL2 film is slightly larger than that of STO substrate. The SAED reveals that $a = 3.905$ Å, and $c_{int} = 3.911$ Å for SL2, which is consistent with the XRD results. The shoulder at the right side of 002 X-ray diffraction reflection, corresponding to the film, is quite clear for $n \geq 6$, but disappears for $n < 4$ (Fig. **S7**), revealing that the out-of-plane lattice constant increases with the decrease of $n$ for SL$n$. Similar enlarged tetragonal ratio around heterostructural interface has been reported in other systems,[39-40] that may affect the electronic structure.

From the above discussions, the enlarged $c_{int}/a > 1$ with breaking of volume conservation around the LSMO/STO interface cannot be accounted by the traditional lattice-match strain effect that would induce $c_{int}/a <1$ under in-plane tensile strain ($a_{LSMO} < a_{STO}$). The crystal and electronic structures are highly correlated in strongly correlated 3$d$ oxide. The electronic charge distribution was examined and verified by Ti $L$ edge electron energy loss spectroscopy EELS,[26,41] summarized in Fig. 2**A**-2**D**. The samples of SL10 and SL6 show the same trend of Ti chemical valence. Taking SL6 as an example, the chemical valence of Ti at LSMO/STO interface (~ 3.78) decreases compared to that at the center of STO layer (~ 3.97), indicating that extra electrons are accepted by the Ti 3$d$ orbitals at the LSMO/STO interface. The averaged chemical valence of Ti is ~ 3.75 in SL2 because each STO unit cell is situated at the LSMO/STO interface. There is no clear trend of chemical valence of Mn[42-43] as revealed in Fig. **S8**. However, a comparison of the shape of the curve (rather than the peak position) of X-ray absorption between simulated and measured Mn $L$ edge XAS suggests a Mn chemical valence higher than bulk value of 3.3 (See **S8** for more details on Mn XAS simulation). Based on the above, the charge transfer from Mn to Ti 3$d$ orbitals is proposed.

The properties of manganite are sensitive to the 3$d$ orbital occupancy, and the Mn $L$ edge X-ray linear dichroism (XLD) was performed to study the electronic configuration, as shown in Fig. 3**A**-3**B**. The difference of two X-ray absorption spectra shows the difference of occupancy in the two $e_g$ orbitals. The gradual decrease in intensity of the feature at ~ 640 eV is due to the gradual increase of relative occupancy of $3d_{3z^2-r^2}$ orbitals with decreasing $n$, as shown in Fig. 3**B**; and SL2 shows the preferred occupancy of $3d_{3z^2-r^2}$ orbitals.[44-45] The resultant depletion of the $3d_{x^2-y^2}$ orbitals weakens the ferromagnetic double-exchange interaction[45] and lowers the paramagnetic-ferromagnetic phase transition temperature in SL2, as shown in Fig. **S4**.

The coupling of electronic structure across the interface has an obvious effect on the angular-dependent magnetoresistance (AMR) at 10 K. In the AMR measurement, the applied magnetic field was rotated from the in-plane to out-of-plane directions. The in-plane current is always perpendicular to the applied magnetic field as shown in the inset of Fig. 3**C**. At 10 K, all films show in-plane magnetic anisotropy (Fig. **S3**). Normally the minimum resistance should occur when the applied magnetic field is in the film plane ($\theta = 90°$). However, with increasing magnetic field from 1 kOe to 80 kOe, the minimum resistance point of SL3 changes from the in-plane to the out-of-plane ($\theta = 0°$) directions (Fig. 3**C**). This phenomenon is related to orbital reconstruction[46] with electronic charge transfer across the LSMO/STO interface, especially the $3d_{3z^2-r^2}$ orbital occupancy. When magnetic field are applied in the film plane, the $3d_{3z^2-r^2}$ band flattens at the Fermi level due to the Mn spin-orbit coupling, and results in a larger resistivity[46] compared to that with magnetic field along the out-of-plane direction due to the decreasing electronic velocity. Fig. 3**D** shows that the AMR effect is largest for SL3 and it decreases with increasing *n*. Although the resistance of SL2 exceeds the low temperature measurement capability as shown in Fig. **S4**, the trend of AMR indicates that the effect of orbital reconstruction at LSMO/STO interface increases with decreasing *n* (enhanced interfacial coupling).

The origin of the preferred occupancy of $3d_{3z^2-r^2}$ at the heterostructure interface had been attributed to arguments such as the interfacial symmetry breaking and new crystal structure phases.[37,44] Based on the existence of the electronic charge transfer and AMR at low temperature, a mechanism due to the orbital hybridization across the interface[47-48] is proposed here (Fig. 4**A**).

From the perspective of molecular orbitals, the interfacial bonding lowers the energy of $3d$ orbitals, favoring the occupancy of the $3d_{3z^2-r^2}$ orbital due to a large spatial overlap. When this interfacial orbital forms, the electronegativity of interfacial bonds may facilitate the charge transfer between Mn and Ti $3d$ orbitals. Due to the correlation between the crystal structure and the electronic structure in $3d$ oxides, the preferred $3d_{3z^2-r^2}$ orbital occupancy may enlarge the out-of-plane $c_{int}$ in order to lower the total energy, as supported by the enlarged $c_{int}/a > 1$ around the left LSMO/STO interface of each LSMO layer (Fig. 2). The oscillating pattern of $c_{int}$ for SL10 and SL6 (Fig. **2A-2B**) indicates a strong coupling across the interface, which may be understood as below: taking one LSMO layer as example, from the left to the right the 1st UC LSMO has strong coupling with its left STO layer, thus it has a large $c_{int}$. The LSMO coupling with STO layer decreases with increasing distance from the interface, and the value of $c_{int}$ of LSMO decreases and reaches a minimum value at the right interface of LSMO layer. The subsequent STO layer is coupled with the preceding LSMO layer. With increasing distance from each LSMO/STO interface, the coupling effect from the LSMO layer decays and the value of $c_{int}$ of STO increases until reaching the next STO/LSMO interface.

The orbital polarization and magnetic anisotropy are not directly correlated[14], but the mechanism of the switching of magnetic anisotropy of superlattice could be addressed based on the electronic hopping integral as below. For a tight-binding Hamiltonian[14] of LSMO ultrathin film, $\sum_R t_{\alpha\beta}(R)e^{iK\cdot R} + \left(\frac{\lambda}{2}\right)\sigma(\theta,\varphi) + \xi L \cdot S$ was proposed to explain the magnetic anisotropy energy (MAE) according to density functional theory. $t_{\alpha\beta}(R)$ represents the hopping integral from orbital α at the original site to orbital β at site **R**. The structural change mainly affects $t_{\alpha\beta}(R)$, which in turn leads to the change of magnetic anisotropy. A high hopping integral is favored along the magnetic easy axis in a ferromagnetically ordered state. In current work, the hopping integral along the in-plane $t_{in}$ and out-of-plane $t_{out}$ direction, illustrated in Fig. **4B**, may shed light on the origin of perpendicular magnetic anisotropy in SL2. The relative intensity of the hopping integral correlates to the detailed shape of the polarization dependent Mn $K$ edge X-ray Absorption Near Edge Structure (XANES)[35, 49-50] as shown in Fig. 4**C-4E** and Fig. **S9**. Upon X-ray absorption, the electronic configuration of MnO$_6$ octahedron in B1 state is proportional to the hopping integral, and then the relative ratio of B1 in polarization dependent XANES is used to characterize the in-plane and out-of-plane hopping integral: the higher intensity ratio ($\Lambda_{//}= I_{B1}/I_{B2}$) in the parallel measurement with the polarization vector ($E$ vector of X-ray) parallel to the film plane suggests a higher out-of-plane hopping integral $t_{out}$.[35, 49-50] For SL2, the out-of-plane hopping integral $t_{out}$ is higher than in-plane hopping integral $t_{in}$ in Fig. **4D**, induced by the enhanced $3d_{3z^2-r^2}$ orbital occupancy, which favors the out-of-plane magnetic anisotropy based on the tight-binding Hamiltonian. In addition, with the decrease of $n$, $\Lambda_{//}$ increases and the $\Lambda_{//}/\Lambda_{\perp}$ ratio also increases as shown in Fig. 4**E**, indicating that the difference of electronic hopping integral along the in-plane and out-of-plane directions increases for decreasing $n$. This is consistent with the trend of $H_i$ with different periodic thickness $n$.

In summary, the variation in periodic thickness of the (La$_{0.67}$Sr$_{0.33}$MnO$_3$)$_n$/(SrTiO$_3$)$_n$ superlattices affects the interfacial atomic arrangement and induced charge transfer and orbital hybridization. Increasing the number of interfaces enhances orbital hybridization, resulting in interfacial electronic charge transfer from Mn $3d$ and Ti $3d$ orbitals and out-of-plane magnetic easy axis. This work demonstrates a promising approach of the use of superlattices to control interface-induced properties of strongly correlated oxides in development of novel magneto-electronic devices.

**Acknowledgements**

The research is supported by the Singapore Ministry of Education Academic Research Fund Tier 2 under the Project No. MOE2015-T2-1-016 and the Singapore National Research Foundation under CRP Award No. NRF-CRP10-2012-02. Work at Brookhaven National Laboratory was supported by the U.S. Department of Energy, Office of Basic Energy Science, Division of Materials Science and Engineering, under Contract No. DESC0012704. PY is supported from SSLS via NUS Core Support C-380-003-003-001. Sector 20 facilities at the Advanced Photon Source, and research at these facilities, are supported by the US Department of Energy - Basic Energy Sciences, the Canadian Light Source and its funding partners, the University of Washington, and the Advanced Photon Source. Use of the Advanced Photon Source, an Office of Science User Facility operated for the U.S. Department of Energy (DOE) Office of Science by Argonne National Laboratory, was supported by the U.S. DOE under Contract No. DE-AC02-06CH11357.

**Captions:**

**Figure 1**: (**A**) Left: Illustration of probability of electronic charge transfer at STO/LSMO interfaces. The probability is normalized to that at interface. The red curve is the probability as electron acceptor, and the blue curve is the probability as electron donor. Right: The illustration of planar charge distribution before (blue bar) and after (pink bar) the charge transfer. (**B**-**C**) the magnetic hysteresis loop at 150 K for SL10 and SL2. The inset in each figure illustrates the effective anisotropy due to the sum of in-plane and out-of-plane magnetic anisotropy. (**D**) The summary of magnetic anisotropy for SL$n$ at 150 K. The pink heart (♥) on the right indicates the magnetic anisotropy of a single LSMO layer with thickness 60 unit cells on SrTiO$_3$ substrate.

**Figure 2**: HAAD-STEM images of (**A-C**) SL10, SL6 and SL2 STO/LSMO superlattice films, respectively. Out-of-plane plane spacing $c_{int}$ (blue squares) and Ti chemical valence (red dots) as a function of position is shown in the middle and bottom part of **A-C**, respectively. The horizontal orange dashed lines mark the $c_{int}$ of STO in bulk substrate. (**D**) Ti EELS at $L_{3,2}$ edges from SrTiO$_3$ substrate far from interface (I: red line), sample surface (II: black line), interface between LSMO/STO superlattice (III: orange circles), center of deposited STO layer in LSMO/STO superlattice (VI: olive circles) in SL6, and deposited STO layer in SL2 film (V: blue circles). The spectra of I and II are close to those of bulk SrTiO$_3$ and LaTiO$_3$, respectively, thus used as the reference spectra of Ti$^{4+}$ and Ti$^{3+}$. Then the fitted Ti valence is Ti$^{3.97+}$, Ti$^{3.78+}$ and Ti$^{3.75+}$ for III, VI and V, respectively. (**E-F**) High resolution TEM (HRTEM) image for SL6 and SL2. The second and third rows are corresponding relative change of out-of-plane (second row) and in-plane (third row) spacing of SrO (La/SrO) planes. The vertical white dashed lines at two STO/LSMO interfaces are for eye guidance.

**Figure 3**: (**A**) The detailed XAS curves for $n = 2$, and the summarized $L$ edge XLD for four samples is shown in (**B**); the parallel measurement (//, red) with electric field of X-ray in the film plane and perpendicular measurement (⊥, blue) with electric field perpendicular to the film plane. (**C**) The effect of magnetic field on the angular dependence of magnetoresistance (AMR) of SL3. All curves are normalized to the resistance at $\theta = 0°$. The measurement configuration is illustrated as the inset of (**C**); (**D**) Summary of AMR at 10 K with magnetic field of 80 kOe for different samples.

**Figure 4**: (**A**) The illustration of the 3$d$ orbital energy levels after hybridization between Mn and Ti across the interface. More discussion in the text. (**B**) The illustration of electronic hopping along the in-plane integral ($t_{in}$) through Mn 3$d_{x^2-y^2}$, and the out-of-plane ($t_{out}$) direction through Mn 3$d_{3z^2-r^2}$. (**C**) The illustration of electronic configuration before and after X-ray absorption. The $L$ denotes the six oxygen ligand around Mn, and $\underline{L}$ denotes one electron missing of the ligand, which locates at Mn site orbital orbitals through hopping between O 2$p_x$ or 2$p_y$ and Mn 3$d_{x^2-y^2}$ orbital. Similar would happen between O 2$p_z$ and Mn 3$d_{3z^2-r^2}$ orbital. See Refs. [36, 49-50] for more details. (**D**) The derivative of the polarization dependent Mn **K** edge XANES with the electric field of the X-ray parallel (//) and perpendicular ($\perp$) to the film plane. There are two features B1 and B2 at the **K** edge. The enhanced B1 peak in the parallel configuration indicates the stronger electronic hopping integral along the out-of-plane direction ($t_{out} > t_{in}$). (**E**) The comparison of XAS for three samples in parallel configuration, which relates to the out-of-plane electronic hopping integral $t_{out}$; The lower inset is the enlarged part around B1 and B2 peaks; the upper inset indicates the trend of $\Lambda_{//}/\Lambda_{\perp}$. $\Lambda_{//}$ and $\Lambda_{\perp}$ indicate the $I_{B1}/I_{B2}$ ratio in parallel and perpendicular configurations, respectively.

**Figure 1:**

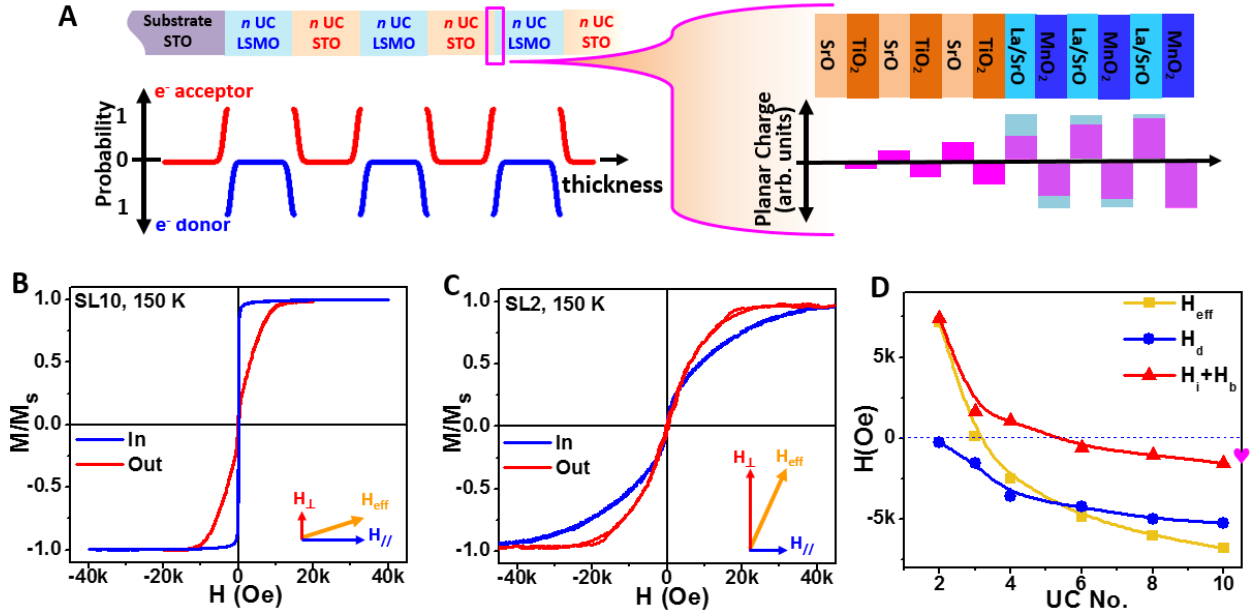

**Figure 2:**

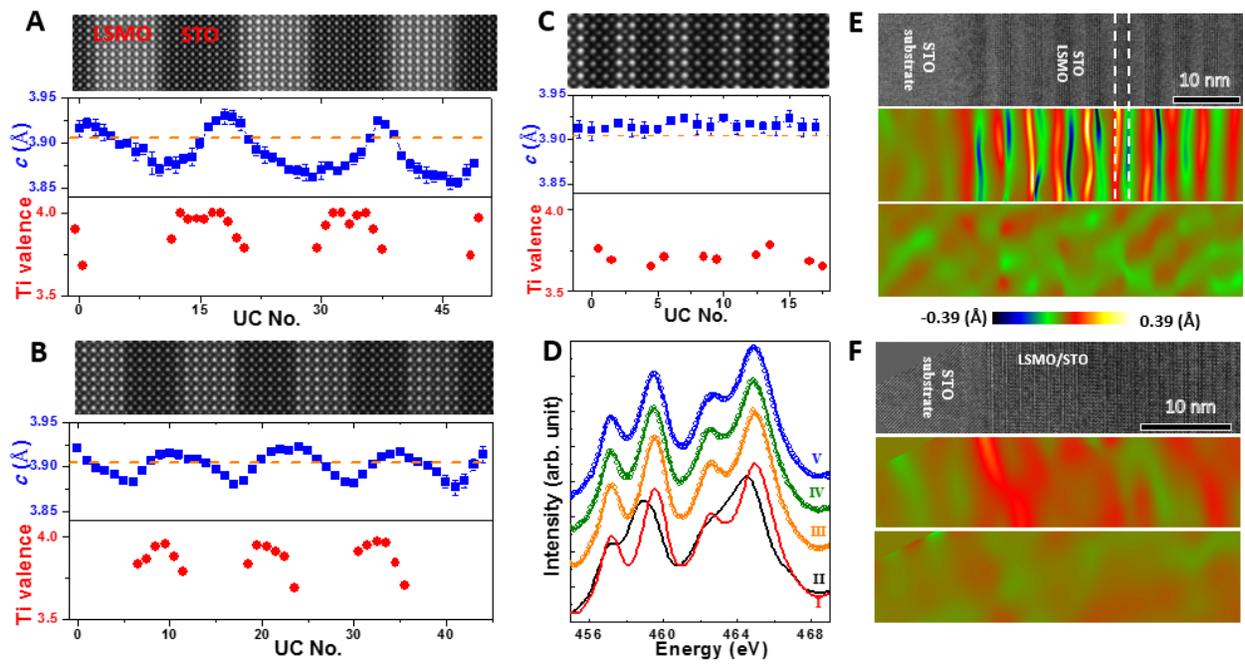

**Figure 3:**

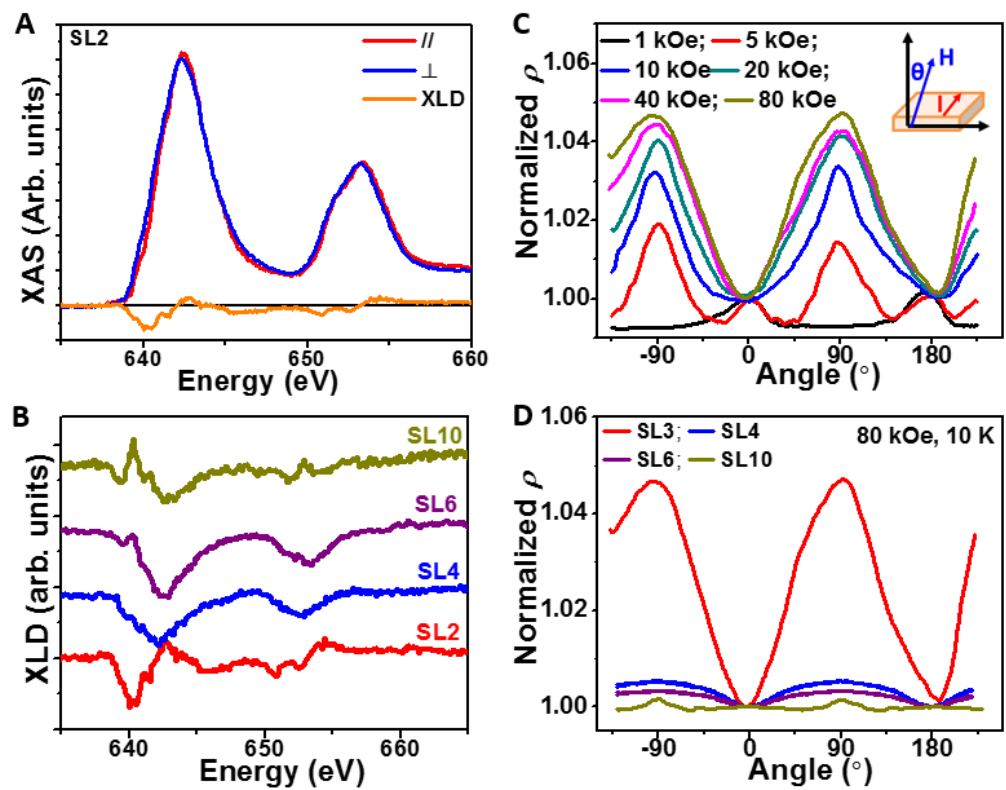

**Figure 4:**

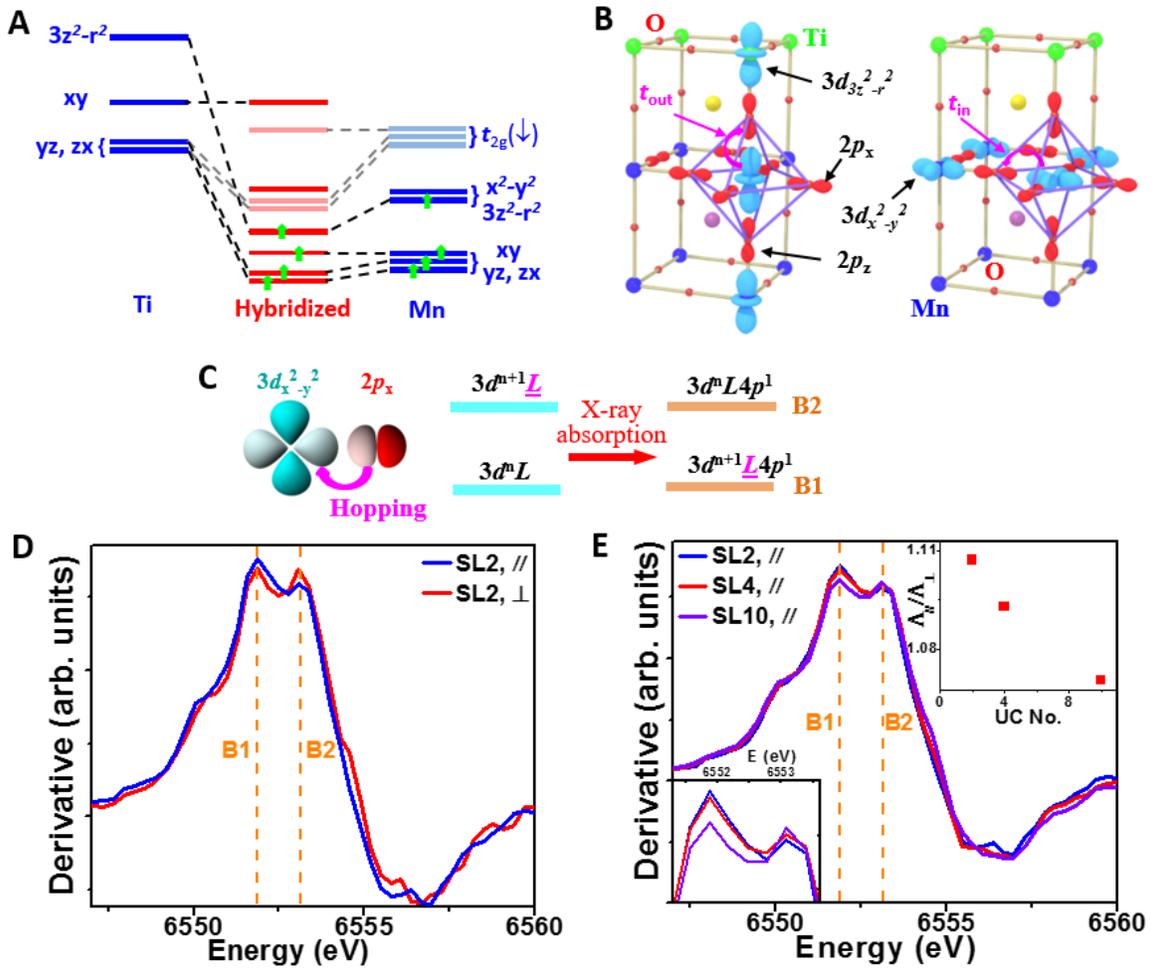

# Supplementary Information
*for*
## Control of magnetic anisotropy by orbital hybridization in $(La_{0.67}Sr_{0.33}MnO_3)_n/(SrTiO_3)_n$ superlattice

Bangmin Zhang,[1,†] Lijun Wu,[2,†] Jincheng Zheng,[2,3] Ping Yang,[4] Xiaojiang Yu,[4] Jun Ding,[1] Steve M. Heald,[5] R. A. Rosenberg,[5] T. Venkatesan,[1,6,7,8,9] Jingsheng Chen,[1] Cheng-Jun Sun,[5,*] Yimei Zhu,[2,*] and Gan Moog Chow[1,*]

[1]Department of Materials Science & Engineering, National University of Singapore, 9 Engineering Drive 1, 117576, Singapore.

[2]Condensed Matter Physics & Materials Science Division, Brookhaven National Laboratory, Upton, New York, 11973, USA.

[3]Department of Physics, Xiamen University, Xiamen 361005, People's Republic of China.

[4]Singapore Synchrotron Light Source (SSLS), National University of Singapore, 5 Research Link, 117603, Singapore.

[5]Advanced Photon Source, Argonne National Laboratory, Argonne, Illinois, 60439, USA.

[6]NUSNNI-Nanocore, National University of Singapore, 2 Engineering Drive 3, 117581, Singapore.

[7]Department of Physics, National University of Singapore, 2 Science Drive 3, 117551, Singapore.

[8]Department of Electrical & Computer Engineering, National University of Singapore, 4 Engineering Drive 3, 117583, Singapore.

[9]NUS Graduate School for Integrative Science and Engineering, National University of Singapore, 28 Medical Drive, 117456, Singapore.

[†]Authors contribute to the work equally.

*Corresponding authors: cjsun@aps.anl.gov, zhu@bnl.gov, msecgm@nus.edu.sg

## Supplemental Materials

**Experimental details**

**S1. Mn *L* edge XMCD**

**S2. Ti *L* edge XMCD**

**S3. Magnetic anisotropy at 10 K**

**S4. Magnetic and transport properties**

**S5. Half-integer diffraction**

**S6. Select area electron diffraction (SAED)**

**S7. X-ray diffraction**

**S8. Mn $L_{3,2}$ edge absorption**

**S9. Polarization dependent Mn K edge XANES**

**Experimental details**

La$_{0.67}$Sr$_{0.33}$MnO$_3$/SrTiO$_3$ films with different thicknesses were grown on (001) SrTiO$_3$ (STO) substrate by pulsed laser deposition equipped with reflection high-energy electron diffraction (RHEED) to monitor the thickness, at 100 mTorr oxygen pressure at a substrate temperature of 780°C. After deposition, the sample was cooled down at 15°C /min in a 200 mTorr oxygen atmosphere. The energy density of the 248 nm KrF excimer laser beam on target was 1.5 J/cm$^2$ at a pulse frequency of 2 Hz.

The magnetic properties were measured by a superconducting quantum interference device (SQUID), and the transport properties were measured by the Physical Property Measurement System (PPMS) using linear four point probe. During measurement, the magnetic field was applied along the in-plane [100] direction.

The crystallographic property of the films at room temperature was studied using a four-circle diffractometer (Huber 4-circle system 90000-0216/0) at the Singapore Synchrotron Light Source (SSLS), with x-ray wavelength equivalent to Cu $K_{\alpha 1}$ radiation.

The X-ray absorption near edge structure (XANES) measurements were carried out using linear polarized X-rays at the undulator beamline 20-ID-B of the Advanced Photon Source (APS), Argonne National Laboratory (ANL), USA. The Si (111) monochromator with resolution δ$E$/$E$ = 1.3 x 10$^{-4}$ was used. The XANES was collected in fluorescence mode using a 12-element Ge solid-state detector. A Mn metal foil, placed to intercept a scattered beam, was used as an online check of the monochromator energy calibration. In order to get reliable information, each XANES curve was obtained by averaging the results of five measurements with total counts more than one million.

The element-specific magnetic information was measured using X-ray magnetic circular dichroism (XMCD). The XMCD at the $L_{3,2}$ edges of Mn were measured with magnetic field of 5 kOe, which were conducted at beamline 4-ID-C at the APS in ANL using total electron mode for data collection at 150 K.

The XANES spectra and XMCD difference spectra were simulated using the FDMNES code.[51] The final states of electron in the cluster with radius of 6 angstrom were calculated using the multiple-scattering approach with spin-orbit coupling included. To account for the core-hole lifetime, the calculated XANES and XMCD spectrum at the $L$ edges of Mn were convoluted with a Lorentzian function. In addition, to account for the experimental resolution and scaling of the spectra, a complementary convolution with a Gaussian function with typical broadening parameters σ=1.0 eV was performed. The spectra of Mn with various chemical valences are normalized for easy comparison.

SAED, HRTEM, STEM-HAADF and EELS spectrum imaging were performed using the double aberration-corrected JEOL-ARM200CF microscope with a cold-field emission gun and operated at 200 kV, at Brookhaven National Laboratory, USA. The microscope is equipped with JEOL and Gatan HAADF detectors for incoherent HAADF (Z-contrast) imaging, Gatan GIF Quantum ER Energy Filter with dual EELS for EELS. The cross-section TEM samples were prepared by focused ion beam (FIB) followed by gentle ion-milling at low operating voltages. The STEM-HAADF images were acquired with condense aperture of 21.2 mrad and collection angle of 67-275 mrad.

## S1. Mn *L* edge XMCD

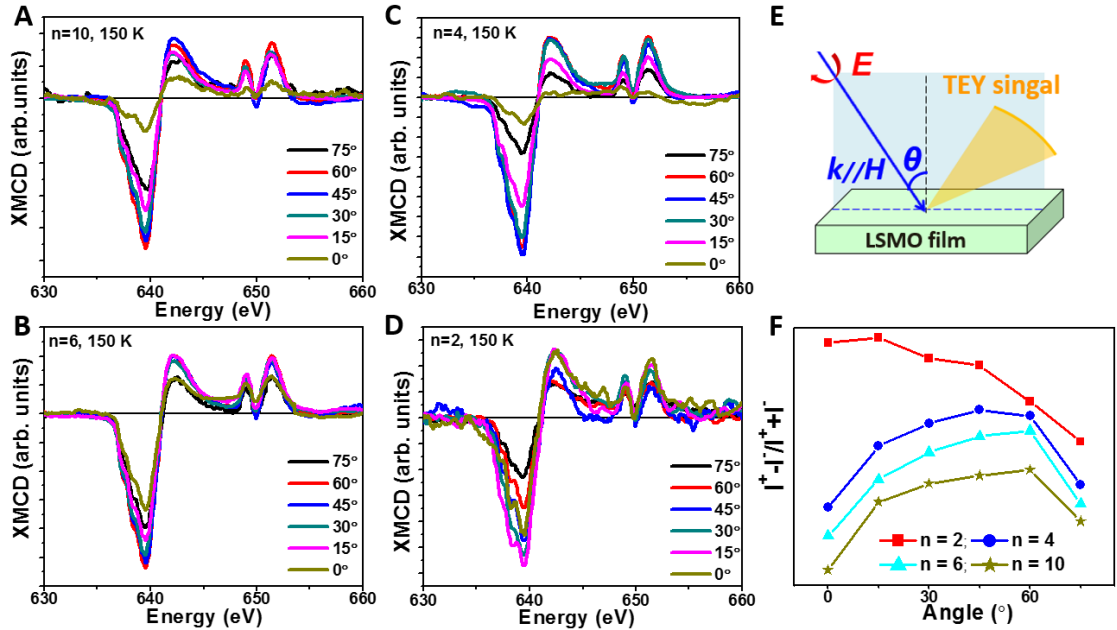

Figure S1: (**A-D**) Angular dependent Mn $L_{3,2}$ edge XMCD for *n* = 2, 4, 6, 10, respectively. The setup of the measurement with different incident angle is illustrated in (**E**). The intensity at peak around 639 eV is summarized in (**F**) to determine the easy axis of the superlattice.

The SQUID was used to provide the overall signal from the whole sample, and the X-ray magnetic circular dichroism (XMCD) yielded element-specific information. Mn $L_{3,2}$ edge XMCD was measured to investigate the magnetic anisotropy for *n* =2, 4, 6, 10, as shown in Fig. S1**A**-S1**D**. The measurement setup is illustrated in Fig. S1**E**. The wave-vector (***k***) of circular X-ray is always parallel to the magnetic field of 5 kOe. The incidence angle $\theta = 0°$ corresponds to the magnetic field normal to the film plane and 90° corresponds to magnetic field in the film plane. The magnetic easy axis is identified by maximum XMCD intensity. Fig. S1**F** shows the summary of XMCD intensity at $L_3$ edge, ~ 639 eV measured at 150 K. When *n* decreases from 10 to 2, the maximum XMCD intensity changes from $\theta = 60°$ for SL10 to $\theta = 15°$ for SL2. Compared to SL10, the easy axis of magnetic Mn ions deviates further away from the in-plane direction in SL2. This is consistent with the results from SQUID measurements, which indicates that the Mn ions have important contribution to the switching of the magnetic anisotropy of SL*n*.

**S2. Ti *L* edge XMCD**

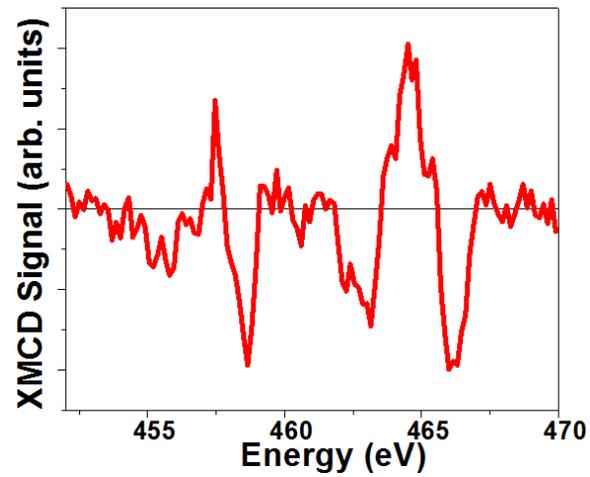

Figure S2: Ti $L_{3,2}$ edge XMCD for SL2 at 150 K with 5 kOe field. According to the sum rules, the estimated magnetic moment of Ti is ~ 0.01 $\mu_B$/Ti. Compared to the moment of Mn, the contribution from Ti is negligible and hence the discussion of magnetic anisotropy of superlattice in this work mainly focuses on the contribution of Mn sites.

**S3. Magnetic anisotropy at 10 K**

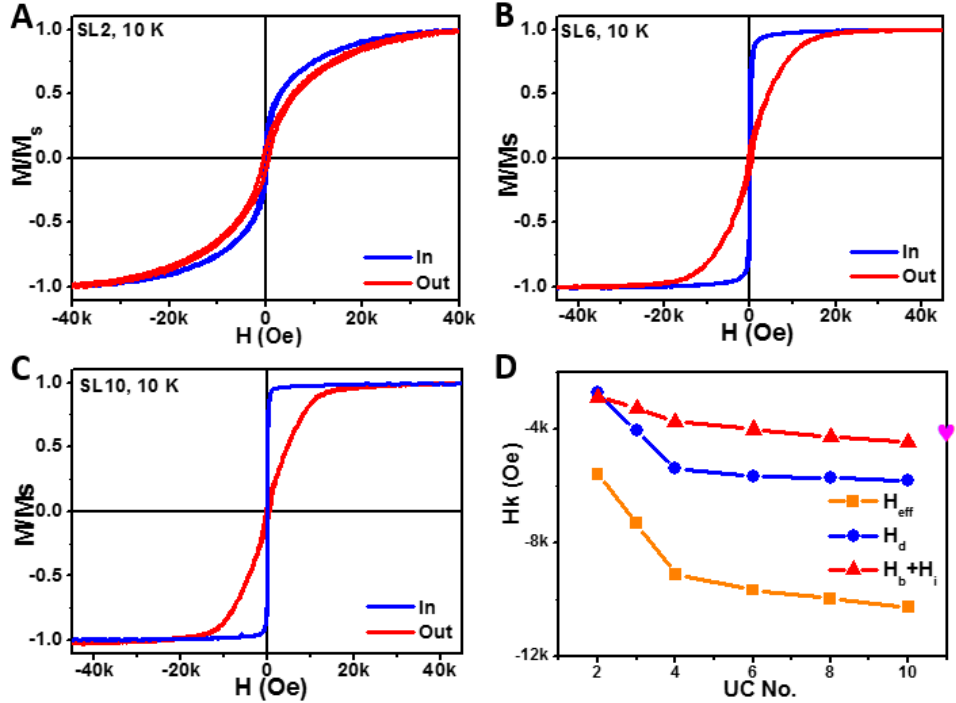

Figure S3: (**A-C**) the MH loop for SL2, SL6, and SL10 at 10 K, respectively; (**D**) the summary of the trend of anisotropy for SL$n$. $H_{eff}$ is the anisotropy calculated from hysteresis loop, $H_d$ is the demagnetization field, and $H_i$ is the interfacial contribution to anisotropy, and $H_b$ is the bulk term which is constant value for all samples. The pink heart (♥) on the right indicates the magnetic anisotropy of a single LSMO layer with thickness 60 unit cells on SrTiO$_3$ substrate.

The magnetic anisotropy at 10 K was measured by SQUID as shown in Fig. S3. The effective anisotropy is separated into constant bulk term $H_b$ and demagnetization term $H_d$ favoring in-plane direction, and interfacial contribution $H_i$ favoring out-of-plane direction. With decreasing $n$, the interfacial contribution $H_i$ to out-of-plane anisotropy increases, and it shows the same trend with varying $n$ at 150 K. Although the easy axis of all SL$n$ still lies in the film plane at 10 K due to the strong demagnetization effect, these results suggest that the effect of interfacial term $H_i$ on magnetic anisotropy with varying periodic thickness exists at a low temperature of 10 K.

## S4. Magnetic and transport properties

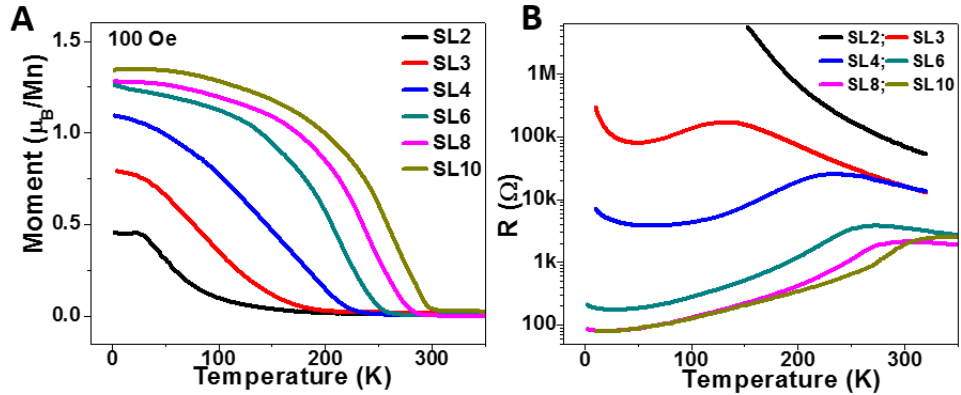

Figure S4: The temperature dependence of (**A**) magnetic and (**B**) transport properties for SL*n*. With decreasing *n*, the Curie temperature decreases and the resistivity increases. See text for more discussion.

## S5. Half-integer diffraction at room temperature

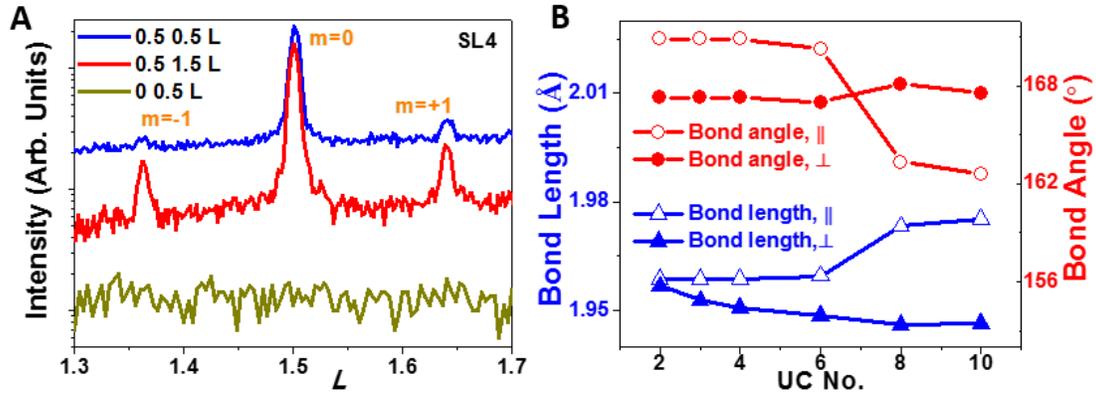

Figure S5: (**A**) Several half-integer diffraction curves for superlattice SL4; (**B**) Summary of bond angle and bond length based on the half-integer diffraction results.

Half-integer diffraction was used to measure the octahedral rotation pattern. The information of bond angle and bond length was obtained by fitting the intensity of half-integer diffraction peaks. Taking SL4 as example, the rotation pattern of $MnO_6$ is $a^-a^-c^-$ according to Glaze's note.[34-36] The sub-peaks, in Fig. S5A, around $L \sim 1.362, 1.638$ are the satellite peaks due to the superlattice configuration. The appearance of these satellite peaks shows the good quality of the sample. The octahedral rotations between neighboring manganite layers are correlated, leading to coherent interference of octahedral rotation. The bond angle and bond length were retrieved by fitting peak intensities using a maple code, and the results are shown in Fig. **S5B**. There is no direct connection between the octahedral rotation and the magnetic anisotropy. Hence, the coupling of crystal structure across the LSMO/STO interface is unlikely the origin of interfacial term $H_i$. See text for more details.

In addition, according to the fitting results, the $MnO_6$ rotations along the two in-plane axes do not change significantly with periodic thickness (4.3° for SL2 and 4.5° for S10), however, the $MnO_6$ rotation along out-of-plane axis ($\gamma$) is suppressed in SL2 ($\gamma = 0.7°$) and increases with the increase of periodic length ($\gamma = 7.5°$ for SL10). The in-plane lattice constant of superlattice is constrained by the STO single crystal substrate, however, the out-of-plane plane spacing exhibits a degree of freedom. Normally the increase of octahedral rotation tends to decrease the lattice constant. Compared to others, SL10 has the increased rotation angle $\gamma$ with shorter out-of-plane plane spacing constant.

## S6. Select area electron diffraction (SAED)

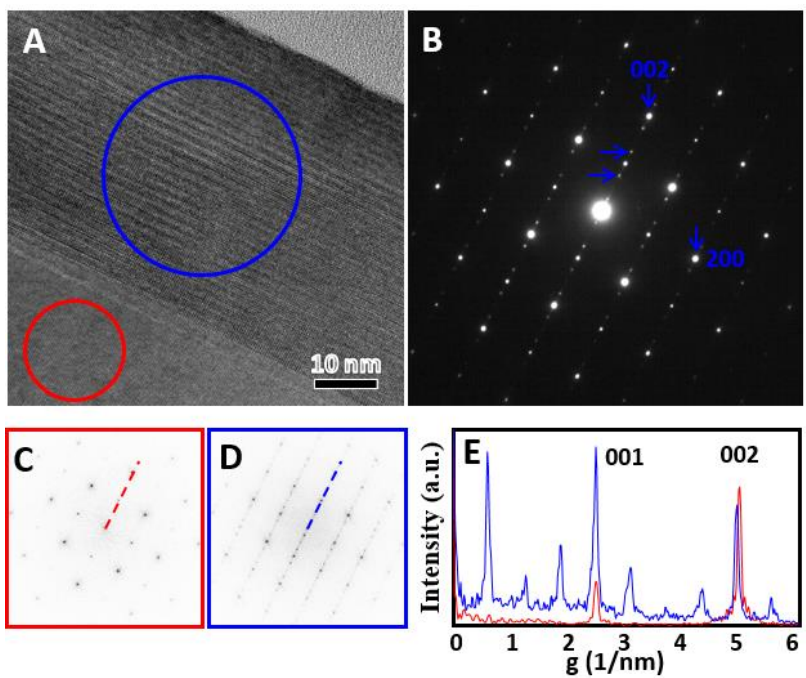

Figure S6: (**A-B**) HRTEM and select area electron diffraction of SL2 film viewed along [010] direction. Sharp superlattice spots (marked by horizontal arrows) corresponding to 4 unit cells LSMO/STO (2 LSMO + 2 STO) present, indicating well grown LSMO/STO superlattice film; Diffractograms from (**C**) red and (**D**) blue circles in (**A**). (**E**) Intensity line profiles from **C** (red dashed) and **D** (blue dashed) scan lines in the diffractograms. (002) peak from **D** shifts to left compared to **C**, indicating increased *c* lattice parameter in the film.

## S7. X-ray diffraction

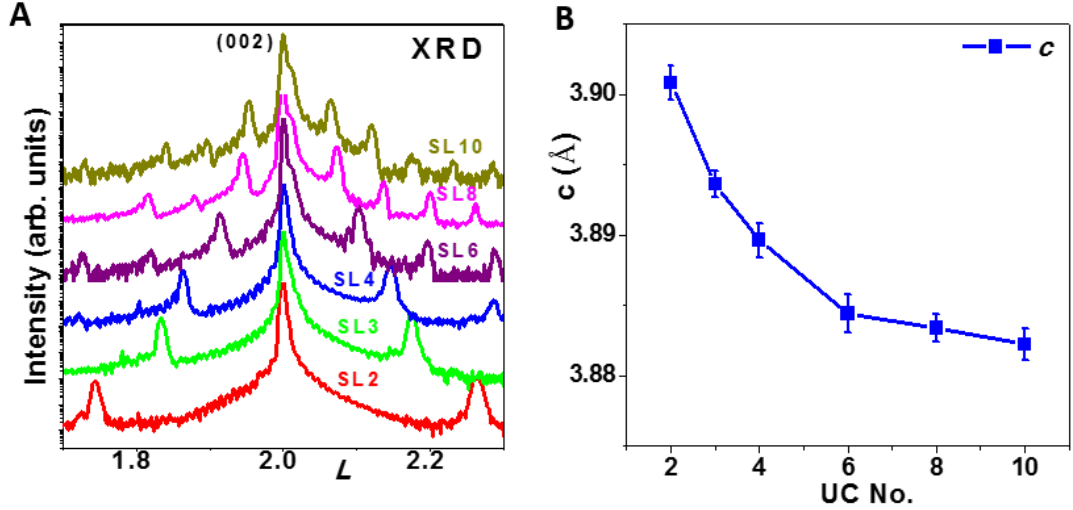

Figure S7: (**A**) X-ray diffraction around 002 peak for all superlative samples measured at room temperature, and (**B**) the summary of calculated out-of-plane lattice constant of SL$n$.

The volume-averaged information using X-ray for superlattice with $n$ = 2, 3, 4, 6, 8 and 10, was measured. In-plane lattice constant $a$ was fixed at 3.905 Å due to the coherent growth of superlattice on STO substrate in all samples, supported by scanning transmission electron microscopy (STEM) image as shown in text. The out-of-plane lattice constant increases with the decreasing $n$, as summarized in Fig. **S7B**. The volume-averaged out-of-plane constant is consistent with the conclusion from the depth profile of lattice constant from STEM.

## S8. Mn $L_{3,2}$ edge absorption

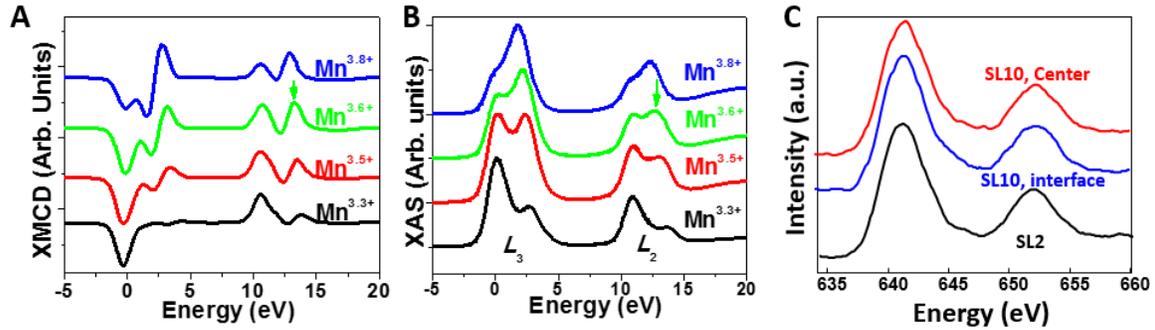

Figure S8: Simulated Mn $L$ edge (**A**) XAS and (**B**) XMCD with different Mn chemical valence. This simulation supports the argument of chemical valence from XAS. (**C**) Mn EELS at $L_{3,2}$ edges from the center (red), LSMO/STO interface (blue) of LSMO layer in SL10, and interface in SL2 (black), respectively.

Simulated Mn $L$ edge XAS was performed to explain the behavior of chemical valence of Mn. Taking SL2 as an example, the calculated Mn $L$ edge XAS changes with varying chemical valence. Due to the competition of the height of two sub-peaks at the $L_3$ and $L_2$ edges, the integrated peak position of $L_3$ and $L_2$ edges with smearing will fluctuate when the valence of Mn varies in the range of (+3.5 ~ +3.8). Hence it is difficult to determine the chemical valence of Mn based on EELS. The curve shape of absorption may yield more information. The simulations indicate that as the Mn valence increases from 3.3 to 3.8, the higher energy shoulder at both the $L_3$ and $L_2$ edges, indicated by the green arrow, increases and dominates for valences greater than 3.6. Compared to the measured XAS (Fig. 3**A**), it suggests that the Mn chemical valence is higher than 3.3. Although the shape of simulated XAS for $Mn^{3.8+}$ is closer to the measured XAS compared to other cases, the simulation may overestimate the interfacial coupling and hence the exact chemical valence in SL is not necessarily 3.8. The simulation on Mn $L$ edge XMCD in Fig. **S8B** supports the same argument.

## S9. Polarization dependent Mn K edge XANES at room temperature

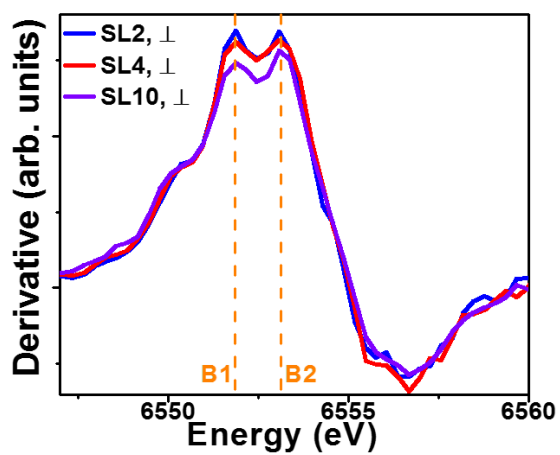

Figure S9: The comparison of XAS for three samples in perpendicular configuration, which relates to the in-plane electronic hopping integral $t_{in}$.